# THE STRENGTH OF GENETIC INTERACTIONS SCALES WEAKLY WITH THE MUTATIONAL EFFECTS

Andrea Velenich [1] and Jeff Gore [1]

[1] Department of Physics, Massachusetts Institute of Technology, 77 Massachusetts Avenue, Cambridge, MA 02139, U.S.A.
Email addresses: velenich@mit.edu , gore@mit.edu

**Abstract:** *Genetic interactions pervade every aspect of biology, from evolutionary theory where they determine the accessibility of evolutionary paths, to medicine where they contribute to complex genetic diseases. Until very recently, studies on epistatic interactions have been based on a handful of mutations, providing at best anecdotal evidence about the frequency and the typical strength of genetic interactions. In this study we analyze the publicly available Data Repository of Yeast Genetic INteractions (DRYGIN), which contains the growth rates of over five million double gene knockout mutants.*

*We discuss a geometric definition of epistasis which reveals a simple and surprisingly weak scaling law for the characteristic strength of genetic interactions as a function of the effects of the mutations being combined. We then utilize this scaling to quantify the roughness of naturally occurring fitness landscapes. Finally, we show how the observed roughness differs from what is predicted by Fisher's geometric model of epistasis and discuss its consequences on the evolutionary dynamics.*

*Although epistatic interactions between specific genes remain largely unpredictable, the statistical properties of an ensemble of interactions can display conspicuous regularities and be described by simple mathematical laws. By exploiting the amount of data produced by modern high-throughput techniques it is now possible to thoroughly test the predictions of theoretical models of genetic interactions and to build informed computational models of evolution on realistic fitness landscapes.*

## Background

Genetic interactions [1] have shaped the evolutionary history of life on earth: they have been found to limit the accessibility of evolutionary paths [2], to confine populations to suboptimal evolutionary states and, on larger time scales, to control the rate of speciation [3]. Epistatic interactions are also relevant to the development of complex human diseases such as diabetes [4]. Complex traits and diseases are influenced by a multiplicity of genomic loci [5], but the independent contributions from each gene often explain only part of the observed phenotype: the phenotype of interest is instead largely determined by the *interactions* among the relevant genes [6].

Despite the broad implications of epistatic interactions, a quantitative characterization of their typical strength is still lacking. In this study we will consider growth rate in yeast as an example of a complex trait modulated by genetic interactions and, supported by a data set containing the growth rates of millions of mutants, consider the following fundamental question: do mutations with larger effects have stronger genetic interactions?

## Results and discussion

### An unbiased definition of genetic interactions

A basic approach to study genetic interactions is to consider two mutations with known effects on a quantitative trait and measure their combined effect in the double mutant [7]. Given [8,9] the growth rates of a wild type *S. cerevisiae* strain ($g_{00}$ = 1) and of two single knockout mutants ($g_{01}$ and $g_{10}$), the growth rate of the double knockout mutant ($g_{11}$) is adequately predicted by a multiplicative null model: $g_{11}/g_{00} = (g_{01}/g_{00})(g_{10}/g_{00})$. Equivalently, defining "log-growth" as the logarithm of the relative growth rate, $G = \log_2(g/g_{00})$, the log-growth of the double knockout mutant is predicted by an additive null model: $G_{11} = G_{01} + G_{10}$ (Fig. 1a).

Epistatic interactions are identified as deviations from the null model, but several non-equivalent alternatives exist for quantifying these deviations [10]. The most common definition of epistasis considers the difference between the measured and the predicted growth rates for the double knockout mutant [8]:

$$e = \frac{g_{11}}{g_{00}} - \frac{g_{01}}{g_{00}}\frac{g_{10}}{g_{00}}$$

Importantly, this definition of *e* subtly constrains the possible values of epistasis. In fact, when combining very deleterious mutations, *e* cannot be large and negative even when the double knockout is a synthetic lethal: $e = 0 - (g_{01}/g_{00})(g_{10}/g_{00}) \sim 0$, if $g_{01} \ll g_{00}$ and $g_{10} \ll g_{00}$. In order to avoid *a priori* constraints on the intensity of epistasis, genetic interactions can be defined as the *ratio* between measured and predicted relative growth rates, leading to:

$$E = \log_2\frac{g_{11}}{g_{00}} - \log_2\frac{g_{01}}{g_{00}} - \log_2\frac{g_{10}}{g_{00}}$$

As an example, $E = +1$ indicates a double mutant whose growth rate is twice as large as would be expected based upon th multiplicative null model, whereas $E = -1$ indicates a double mutant whose growth rate is half as large as predicted. This definition of epistasis as fold deviation in the multiplicative model for growth rates is equivalent to a natural definition of epistasis as linear deviation in the additive model for log-growth rates (Fig. 1b):

$$E = (G_{00} + G_{11}) - (G_{01} + G_{10})$$

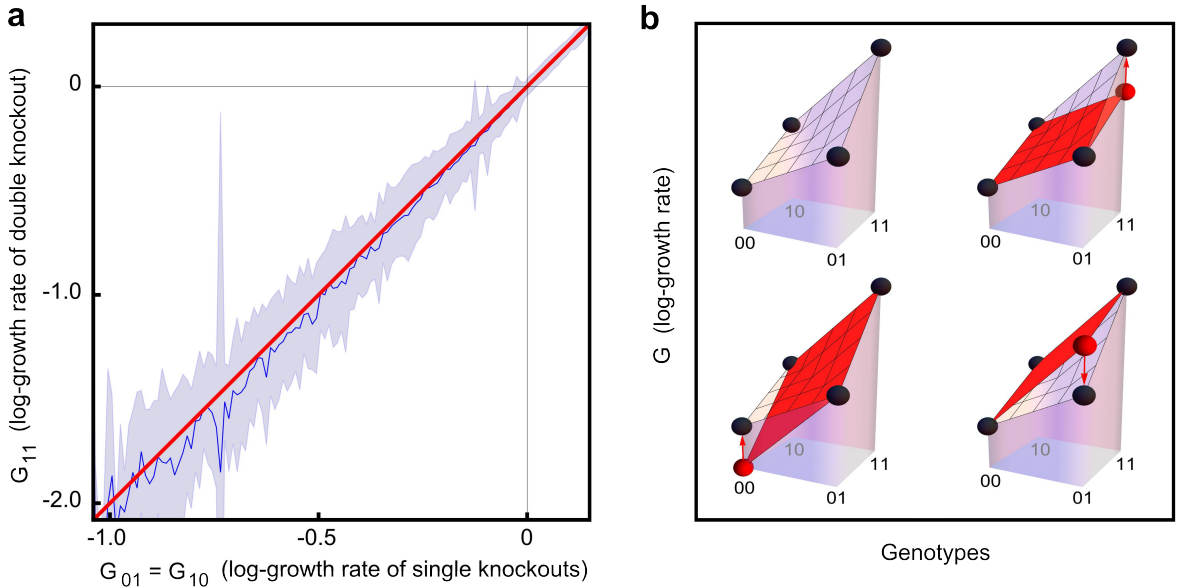

**Figure 1 | The log-growth rates of two mutations combine additively. (a)** The average effect of a double knockout ($G_{11}$) as a function of the effects of the single knockouts ($G_{01}$ and $G_{10}$) is $G_{11} = G_{01} + G_{10}$. Experimental mean +/- standard deviation (blue line and blue shaded area) and prediction of the additive null model (red line). **(b)** Given two mutations there are four possible mutants with their corresponding log-growth rates (black dots). If three of the four log-growth rates are known, the fourth one can be predicted by a linear extrapolation (red plane) and epistasis can be defined as the linear deviation from such prediction (red arrow). The magnitude of the deviation is the same regardless of which three of four mutants are chosen.

A second bias of the common definition of epistasis is that *e* depends on the choice of which genotype is labeled as "wild type" or "00", a choice which is always arbitrary, but more obviously so when studying engineered organisms or populations evolving in alternating environments [11]. On the contrary, $|E| = |(G_{00} + G_{11}) - (G_{01} + G_{10})|$ depends only on which pair of genes is considered, being a geometric measure for the "curvature" of the fitness landscape (Fig. 1b).

The definition of $E$ found some favor in the theoretical literature [12,13], but it is not routinely used to analyze experimental data apart from rare exceptions [14,15]. Its main drawback is that synthetic lethals have a log-growth rate of minus infinity and require a separate although simpler analysis in which lethal interactions can simply be counted. The definition of $E$ proves instead to be extremely valuable when quantifying the *strength* of non-lethal genetic interactions.

**Epistatic interactions scale weakly with the mutational effects**

With the appropriate definition of epistasis, a simple relation between the growth rate effects of two mutations and the expected strength of their interaction emerges.

Let us consider two groups of mutations: in the first group all mutations have log-growth effect $G_{01}$ and in the second group all mutations have log-growth effect $G_{10}$. We can then build all possible double mutants obtained by combining one mutation from each group. In the absence of epistasis, all the double mutants have a log-growth rate $G_{11} = G_{01} + G_{10}$ and the distribution of genetic interactions is a delta function. When epistasis is present, the distribution of genetic interactions has a finite standard deviation $\sigma_{G01,G10}$, which can be used as a quantitative indicator of the characteristic strength of epistatic interactions between two mutations whose effects are $G_{01}$ and $G_{10}$. Importantly, the distribution of genetic interactions is still concentrated around zero, since the null model remains approximately valid (Fig. 1a and Fig. 2d).

In order to produce reliable numerical results, thousands of growth rates are necessary to characterize the probability distribution of epistasis. Previous studies [13, 15-17] on the relation between the growth effects of a mutation and its epistatic interactions have often been based on a handful of mutations and only in recent years anecdotal evidence is being replaced by robust statements based on large data sets. Perhaps the most impressive of these data sets is the one made publicly available [8] by the DRYGIN collaboration. The genome of the budding yeast *S. cerevisiae* includes approximately 6,000 genes, about 1,000 of which are essential. Viable mutants can be constructed by knocking out any of the ~5,000 non-essential genes, by reducing the expression of the essential genes, or by partially compromising the functionality of their gene products. The DRYGIN data set (Fig. S1) has been compiled with the growth rates of about 5.4 million double knockout mutants, a sizable fraction of all possible double knockout mutants in yeast.

We analyzed the DRYGIN data set by binning pairs of mutations according to the log-growth effects of their single knockouts $G_{01}$ and $G_{10}$, following the method described above to outline the probability distribution of epistasis. We chose bin sizes which grow exponentially with G to ensure an approximately constant number of data points in each bin (Fig. S2). Most bins contain from thousands to tens of thousands data of points. For each bin we computed the variance in the epistatic interactions, var($G_{01}$, $G_{10}$) = $\sigma^2_{G01,G10}$ (Fig. 2a). The square root of that variance represents then the strength of epistasis as a function of the independently varying effects of the two single knockouts. A natural expectation for the dependence of epistasis on the effect of the combined mutations comes from rescaling Figure 1a: if all the log-growth effects of single and double knockouts increase by a factor of two, then also the strength of epistasis should increases by a factor of two. Surprisingly, when combining *deleterious* mutations, the strength of epistatic interactions does grow with the effects of the mutations that are combined, but the dependence is much weaker: when the effect of both single knockouts doubles, the strength of epistasis increases only by a factor of Sqrt(2) (Fig 2).

In more detail, we observed that if the effect of the first knockout ($G_{01}$) is held constant, the dependence of the variance of epistasis on the effect of the second knockout ($G_{10}$) is well approximated by a Michaelis-Menten law (Fig. 2b):

$$\mathrm{var}(G_{10}) = v \frac{|G_{10}|}{K + |G_{10}|}$$

When both knockouts' effects are free to vary, the requirement that the variance is a symmetric function of its two variables, $G_{01}$ and $G_{10}$, implies that K = $|G_{01}|$ and that v is proportional to $G_{01}$. A one-parameter function which fits the observed variance over the whole range of deleterious fitness effects (Fig. 2a) is then:

$$\mathrm{var}(G_{01,}G_{10}) = 2c\frac{|G_{01}||G_{10}|}{|G_{01}| + |G_{10}|}$$

with c = 0.079. This functional form can also be obtained from a simple model based on diffusion in fitness space (Supplementary Text 1). An even simpler phenomenological fit, although slightly less accurate, is: $\text{var}(G_{01}, G_{10}) = c\ \text{Sqrt}(|G_{01}||G_{10}|)$ (Fig. S3). Importantly, these functions capture two major features of the data: first, epistasis vanishes when $G_{01}$ or $G_{10}$ are zero; second, when the effects of the two knockouts are similar ($G_{01} = G_{10} = G$ along the diagonal of the surface in Figure 2a), the variance of epistasis is approximately proportional to G: $\text{var}(G_{01}, G_{10}) = c\ |G|$ (Fig. 2c).

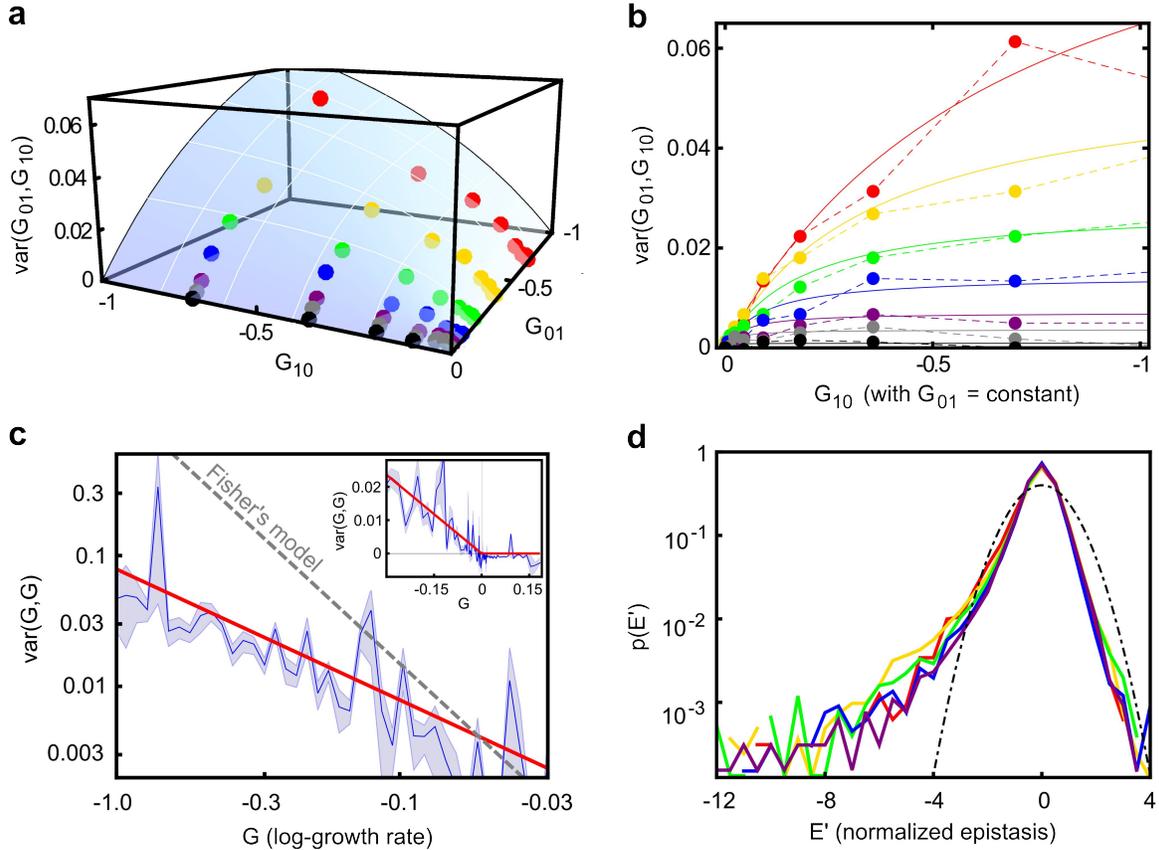

**Figure 2** | **The strength of epistatic interactions scales with the log-growth effects of the interacting knockouts. (a)** Each dot represents the variance of several thousand epistatic interactions binned according to the log-growth effects of the two single knockouts, $G_{01}$ and $G_{10}$. The blue surface is the phenomenological fit: $\text{var}(G_{01}, G_{10}) = 0.079 \times 2\ |G_{01}||G_{10}|\ /\ (|G_{01}| + |G_{10}|)$. **(b)** Slices of the plot in (a) for $G_{01}$ = constant. The dots are the same as in (a) and the solid lines represent the corresponding slice of the one-parameter fitting surface. **(c)** Diagonal slice of the plot in (a) with finer bins ($G_{01} = G_{10}$ within 20%, G = mean($G_{01}, G_{10}$)). The blue shaded area is the 25%-75% confidence interval computed by bootstrap; the red line ($\text{var}(G, G) = 0.079\ G$) is computed from the phenomenological model and the dashed gray line, for which $\text{var}(G, G)$ is proportional to $G^2$, represents the lower bound to the slope predicted by the Fisher's geometric model. **(c, Inset)** The epistatic interactions between beneficial mutations are vanishingly small, independently of the effect of the combined mutations. **(d)** Probability density functions p(E') for the strength of genetic interactions between two deleterious knockouts with similar log-growth effects. Different colors correspond to knockouts with different effects: the growth rates effects of the single knockouts being combined are close to -38% (red), -22% (yellow), -12% (green), -6% (blue), -3% (purple). Each curve has been rescaled so that all distributions have standard deviation one. The left tail of the distributions displays a fat tail, describing the occurrence of strong negative genetic interactions (the dashed-dotted black line is a normal distribution, for comparison).

The scaling described above is observed only for deleterious knockouts. When combining the beneficial knockouts in the DRYGIN data set, instead, the strength of epistasis is close to zero (Fig. 2c, inset). This might be because the slightly beneficial knockouts are not adaptive mutations, but simply remove genes which are not needed in the conditions chosen for the experiment, so that their interactions are likely to be negligible. In apparent contrast with this observation, recent studies [15,18] on adaptive mutations in *E. coli* suggest that genetic interactions between adaptive mutations are mostly negative. In fact, during adaptation the prevalence of negative interactions is likely to be caused by biased sampling, since the mutations which fix in the population are likely the ones that solve environmental or biological challenges for an organism. Diminishing returns arise because the appearance of multiple "solutions" to the same challenge is not necessarily preferable over the presence of a single solution. Rather than focusing on mutations that fix during a bout of adaptation, the DRYGIN data set includes a large fraction of all possible pairs of genes in the yeast genome. Since for most pairs the two genes are involved in unrelated biological processes, interactions are often vanishingly small. We did observe, however, that the distribution of epistatic interactions is asymmetric, with a heavy tail of deleterious interactions (Fig. 2d).

**Comparison between theory and experiment**

The scaling of epistasis observed in the DRYGIN data set is in sharp contrast with the predictions of Fisher's geometric model [19], a popular model of epistasis in which genetic interactions emerge from geometry. As we saw, when the effects of the two knockouts are similar ($G_{01} = G_{10} = G$), the variance of epistasis is approximately proportional to G. In the Fisher's model, instead, the variance var(G, G) would grow faster than $G^2$ (Fig. 2c and Supplementary Text 2), a much stronger dependence than the linear dependence observed experimentally.

A concrete numerical example can highlight the importance of the weaker-than-expected scaling of epistasis described in this study. Let us consider two gene knockouts, each of which reduces the relative growth rate by 5%, from 1.0 to 0.95. According to the multiplicative null model, the growth rate of the double knockout will be approximately $0.95^2 \sim 0.90$. The question is now: "What kind of deviations could be expected around 0.90? Would a growth rate of 0.85 be surprising? What about a growth rate of 0.50?". Let us use the analytic fit discussed in the previous section: since $g_{01} = g_{10} = 0.95$, then $G_{01} = G_{10} = \text{Log}[2, 0.95] = -0.074$, $G_{11} = G_{01} + G_{10} = -0.148$ and $\sigma(G_{01}, G_{10}) = 0.076$. A +/- 1-sigma interval for the growth rate of the double knockout is then $[2^{-0.148 - 0.076}, 2^{-0.148 + 0.076}] = [0.86, 0.95]$. Notice how it is not unlikely for epistasis to cancel the effect of the second mutation, so that the growth rate of the double knockout mutant is higher than 0.95, the growth rate of either of the single knockout mutants.

Let us now consider two gene knockouts with stronger effects, each of which reduces the growth rate from 1.0 to 0.60. Then $G_{01} = G_{10} = \text{Log}[2, 0.60] = -0.737$, about ten times as large as the log-growth of the single mutants in the previous example. The Fisher's model would predict a sigma(G, G) at least ten times larger than in the previous example ($\sigma(G, G) \geq 0.76$), and an interval of likely growth rates for the double knockout mutants at least as large as $[2^{-1.47 - 0.76}, 2^{-1.47 + 0.76}] = [0.213, 0.610]$. Notice how, once again, it is not unlikely that due to genetic interactions the growth rate of the double knockout mutant is higher than 0.60, the growth rate of either of the two single knockout mutants. The analytic model derived from the experimental data leads to a strikingly different conclusion: $\sigma(G_{01}, G_{10}) = 0.241$, and the +/- 1-sigma interval for the growth rate of the double knockout becomes $[2^{-1.47 - 0.241}, 2^{-1.47 + 0.241}] = [0.305, 0.425]$. In this case a deviation from the null model larger than three standard deviations would be needed for the double knockout mutant to have a growth rate higher than the single knockouts' growth rate of 0.60, making the event extremely unlikely.

**Epistasis constrains the evolutionary dynamics**

The last section provided two examples of reciprocal sign epistasis, realized when two deleterious mutations produce a double mutant fitter than either of the two single mutants (Fig. 3a). In those cases, a fitness valley limits the evolutionary accessibility of the fitter double mutant, and only on longer time-scales the simultaneous appearance of two mutations [20,21] may drive a population to the new local fitness maximum. In this context the scaling behavior of epistasis is of great importance, since it determines the number and the topology of the evolutionarily accessible paths [2,22,23], ultimately affecting the possible outcomes of the evolutionary process.

In order to describe how epistasis shapes naturally occurring fitness landscapes, let us consider S(G, G), the probability to observe sign epistasis when combining two mutations with similar growth rate effects G. S(G, G) depends on the typical interaction strength, $\sigma_{G,G}$ = Sqrt(var(G, G)). In particular, if $\sigma_{G,G}$ is proportional to G, then the probability to observe sign epistasis is independent of G. The Fisher's model implies a super-linear dependence of $\sigma_{G,G}$ on G, thus predicting a higher probability to observe sign epistasis among mutations with strong effects. Instead, the scaling of $\sigma_{G,G}$ observed in the DRYGIN data set is proportional to Sqrt(G) implying that sign epistasis is less likely to occur among mutations with large effects (Fig. 3b). A direct consequence of the scaling of epistasis described here is a roughening of the local fitness landscape in proximity of an evolutionary optimum: when the fitness effects of available mutations become small [24], epistatic interactions become increasingly relevant [25,26], reducing the accessibility of evolutionary paths and further slowing down the rate of adaptation [27,28]. The evolutionary dynamics on correlated fitness landscapes [17,29] with the realistic correlations described here certainly deserves further theoretical investigation.

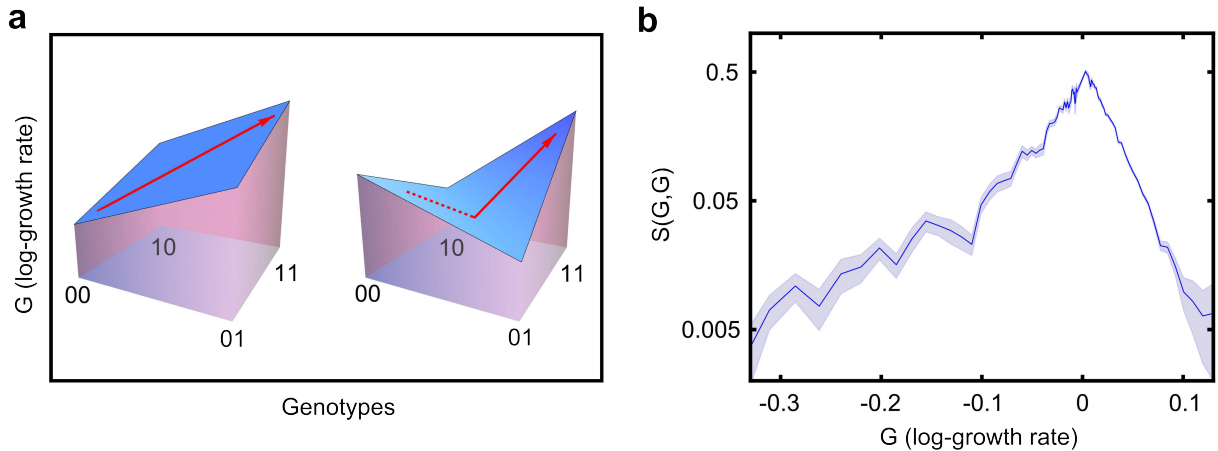

**Figure 3 | Sign epistasis is less likely to occur between mutations with large effects. (a) Examples of a smooth landscape with paths of monotonically increasing fitness (left) and a rugged landscape characterized by reciprocal sign epistasis (right). (b) Experimentally measured probability of observing sign epistasis as a function of the log-growth of two single knockouts with similar effects ($G_{01}$ = $G_{10}$ within 20%, G = mean($G_{01}$, $G_{10}$)). The blue shaded area is the s.e.m. computed by bootstrap.**

**Experimental uncertainty generates spurious epistatic interactions**

When inferring genetic interactions from experimental data, it is important to consider that each measured growth rate is affected by some uncertainty and that measurement errors in the growth rates could erroneously be interpreted as genetic interactions. Importantly, for each single and double mutant, the DRYGIN data set provides the mean growth rate together with its estimated experimental uncertainty (the growth rate of each mutant being measured at least four times).

In order to quantify the effect of the experimental uncertainty on the inferred epistatic interactions, we constructed a number of mock data sets, assuming that the null model without epistatic interactions described biology exactly: in these data sets, each single knockout had the same growth rate as in the original DRYGIN data set and each double knockout had a growth rate equal to the product of the relative growth rates of the corresponding single knockouts. We then randomized the mock data sets by shifting each growth rate by a random amount sampled from a student's t-distribution with width depending on the corresponding experimental uncertainty reported in the DRYGIN data set (Supplementary Text 3). As expected, the analysis of these "noisy" data sets revealed some epistasis, clearly caused by our addition of experimental noise rather than by any biological mechanism. We found that for pairs involving beneficial or neutral mutations, the variance computed in the mock data sets was comparable or even higher than the variance observed in the DRYGIN data set (e.g. Fig. 4a black curves and Fig. 4b blue regions). This fact provides an important internal control, suggesting that the experimental noise has not been underestimated. In spite of this, for pairs of knockouts with substantially deleterious effects, experimental noise accounted for less than half of the total observed variance, the rest representing genuine biological interactions (Fig 4a red curves and Fig. 4b red regions).

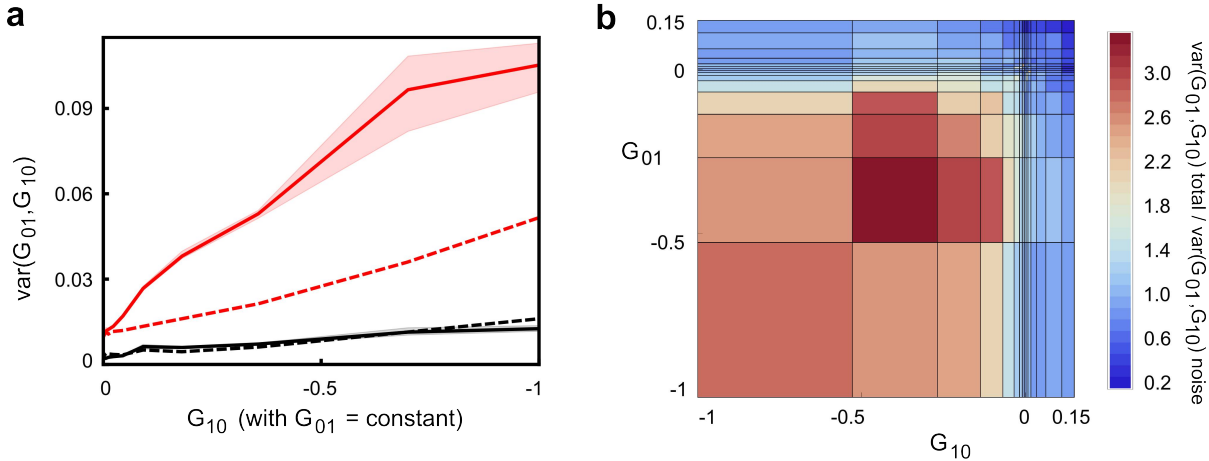

**Figure 4 | Experimental noise does not account for all of the observed variance of epistasis. (a) Comparison of experimentally measured variance (solid lines; shaded areas: 25%-75% confidence intervals) and variance caused by experimental noise (dashed lines). If one of the two mutations is neutral, noise accounts for all of the observed variance (black). When deleterious mutations are combined noise accounts for less than half of the observed variance (red, $G_{01} \sim -0.7$). (b) Ratio between total observed variance and noise-generated variance as a function of the log-growth of the knockouts being combined. For deleterious knockouts the ratio can be significantly larger than one.**

We then decomposed the variance observed in the original DRYGIN data set into a contribution due to experimental uncertainty and a contribution of biological origin; the strength of epistatic interactions was finally computed as the square root of the biological part of the variance. For deleterious knockouts, the relative difference between epistasis computed from the raw data and from the data after subtracting the experimental noise was less than 30%, emphasizing the significant but not overwhelming contribution of experimental noise to the observed variability. Figures 2a, 2b and 2c represent the "biological" part of the observed epistasis; before subtracting the contribution of the experimental uncertainty the plots are qualitatively similar, but quantitatively slightly different (Fig. S4). Importantly, since variances are additive, the estimated contribution of the experimental uncertainty to epistasis is largely independent of the choice of the statistical distribution used to model experimental uncertainty. In two instances, however, the unknown details of the full distribution of experimental noise are important: when outlining the distribution of epistatic interactions (Fig. 2d) and when describing the probability to observe sign epistasis (Fig. 3b). In these last two figures we plotted the raw data and did not attempt to de-convolve the contribution of experimental uncertainty.

**The scaling of genetic interactions may be generic**

Our analysis has so far been limited to interactions between entire gene knockouts. Although mutations with extreme effects on gene regulation and horizontal gene transfer are biologically relevant mechanisms for the removal or the acquisition of whole genes at once, organisms explore possible genetic variants largely through the accumulation of single point mutations. The DRYGIN data set contains thousands of double mutants for which the first mutation is a gene knockout while the second mutation consists of one or more point mutations in a different gene, causing the gene product to misfold in a temperature sensitive way. Although the distribution of growth rate effects for point mutations is different than for single gene knockouts (Fig. S2), the statistics of genetic *interactions* is remarkably similar when combining two single knockouts and when combining a single knockout with a point mutation (Fig. 5). A similar scaling is observed also for the epistatic interactions between single gene knockouts and DamP (Decreased Abundance by mRNA Perturbation [30]) perturbations of a second gene (Fig. S5). The analysis of these hybrid double mutants suggests that the statistics of the interactions between any two genetic perturbations is determined only by their growth rate effects [31] and not by their biological origin in terms of point mutations or gene knockouts.

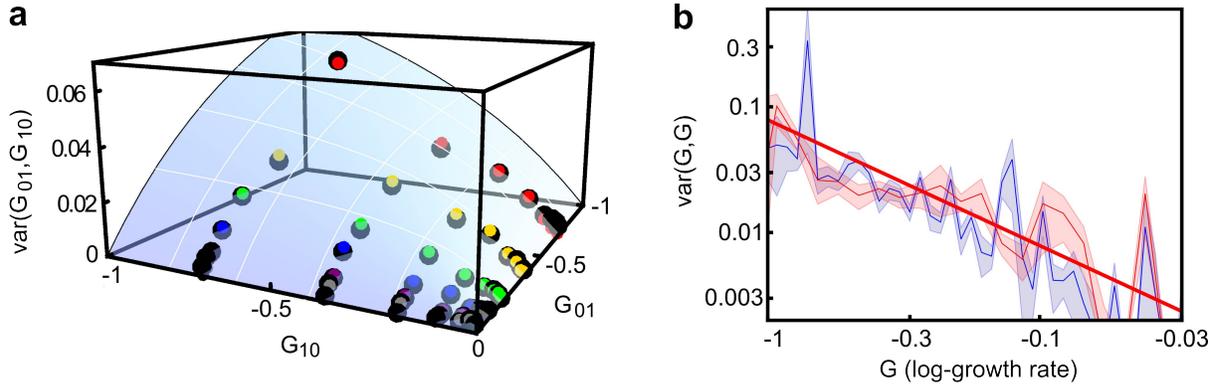

**Figure 5** | **Point mutations have similar epistatic interactions as entire gene knockouts. (a) Comparison between the variance observed in double gene knockout mutants (rainbow dots, same as in Fig. 2a) and the variance observed in mixed double mutants generated by combining a gene knockout and point mutations in a different gene (black dots). (b) The red curve is the diagonal slice of the plot in (a) ($G_{01} = G_{10}$ within 20%, $G$ = mean($G_{01}$, $G_{10}$)) and the red shaded area is the 25%-75% confidence interval for the mixed double mutants variance. For comparison it is superimposed to Figure 2c describing the variance for double gene knockouts (blue). As in Figure 2c the red line has equation var($G$, $G$) = 0.079 $G$.**

**A comparison between different definitions of epistasis**

Importantly, any quantitative result on epistasis is a consequence of how epistasis is defined . As an example, if the DRYGIN data set is analyzed using the "traditional" definition of genetic interactions $e = g_{11}/g_{00} - (g_{01}/g_{00}) (g_{10}/g_{00})$, then the linear dependence of var($G$, $G$) on $G$ in Fig. 2c is replaced by an oddly non-monotonic dependence, displaying weaker interactions for pairs of genes with either very small or very large fitness effects (Fig. 6a). As mentioned previously, this decrease in the inferred strength of epistatic interactions for very deleterious mutations is a mathematical consequence of the traditional definition of epistasis, not a property of genetic interactions. The same bias would lead us to conclude that genes with strong effects on growth are almost non-interacting (Fig 6b, red line). However, since previous studies have determined that essential genes partake in more interactions than non-essential genes [32], it is also reasonable to expect that non-lethal genes with large growth effects are involved in more interactions than genes with small growth effects. Indeed, according to the "geometric" definition of epistasis $E = G_{11} - (G_{01} + G_{10})$, the fraction of genes with which a gene interacts steadily increases with the growth rate effect of the gene (Fig. 6b, blue line). The traditional definition of epistasis consistently overlooks interactions between genes with large growth rate defects, as confirmed by a further analysis comparing the two definitions of epistasis against interactions inferred from the Gene Ontology database [33] (Fig. S6). According to the geometric definition of epistasis, genetic networks [34] are denser than expected not only among essential gene [32], but also among genes with large growth effects.

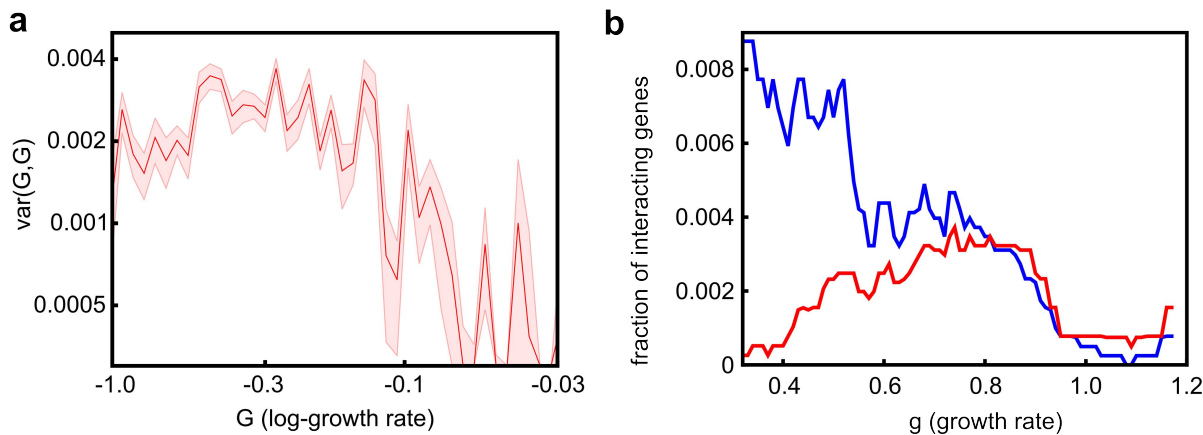

**Figure 6 | Comparison between the traditional and the geometric definitions of epistasis (*e* and *E*, respectively). (a)** Figure equivalent to 2c, using the traditional definition of epistasis. **(b)** The fraction of genes interacting with a given gene is a function the growth rate effect of such gene. Only the 10,000 most interacting pairs according to the geometric definition (blue) and the traditional definition (red) are considered as "interactions".

## Conclusions

We analyzed the growth rates of about five million double mutants in the DRYGIN data set and characterized how the strength of genetic interactions depends on the growth effects of the mutations being combined. We found a weaker dependence than what is predicted by current theoretical models and, although the results have been obtained mainly from entire gene knockouts, there is some evidence that the observed scaling might extend to the interactions between single point mutations. The scaling of epistasis might or might not be generic [35,36]: important drivers could be the harshness of the environment [37], details about the evolutionary history [38-40], sexual vs. asexual reproduction [41] and, perhaps most importantly, metabolic [42-45] and genetic complexity [46,47]. In general, the experimentally observed scaling suggests a previously unexplored class of correlated fitness landscapes with tunable roughness, in which epistasis depends explicitly on the effects of the mutations being combined.

A clear limitation of our discussion is that only pair interactions have been considered. While high-throughput experiments will provide data on higher-order interactions, a solid understanding of pair interactions remains necessary before addressing n-mutation interactions. A genuine 3-mutation interaction, for instance, should be defined as the unexplained deviation from what can be computed by combining the effects of all relevant mutations and their pair interactions [17,48], perhaps using linear fits within the additive null model for log-growth rates.

The results we presented are based on a geometric definition of epistasis. We compared this definition with a more standard definition, highlighting the desirable mathematical properties of the geometric definition and the simple phenomenological relations it produces.

In conclusion, although each epistatic interaction between specific genes depends on biological details and remains largely unpredictable from first principles, we have shown that the statistical properties of an ensemble of interactions can display conspicuous regularities and be described by simple mathematical laws.

## Materials and methods

The DRYGIN data set is publicly available at http://drygin.ccbr.utoronto.ca/~costanzo2009/.
The data base file sgadata_costanzo2009_rawdata_101120.txt.gz was downloaded on 08/17/2010 and analyzed with Mathematica (code available upon request). We restricted our analysis to double knockouts mutants whose growth rates were positive numerical values and for which the growth rates of both single mutants were numerical values (see Supplementary Figure 1). Some genes appear in the data set both as query and array genes; care has been taken to avoid double counting.

In order to quantify the contribution of experimental uncertainty to epistasis we generated nine randomized mock data sets. The mean level of noise-generated epistasis on these nine data sets is reported in Figure 4 (dashed lines). See Supplementary Text 3 for an extensive discussion of the choice of student's t-distributions to generate the mock data sets from the original data set.

The file go_201207-assocdb-tables.tar.gz for the mySQL Gene Ontology database was downloaded on 07/29/2012 from http://www.geneontology.org/GO.downloads.database.shtm. The database was queried with Python and analyzed Mathematica (code available upon request).

**Acknowledgments**

We are grateful to Mingjie Dai for collaboration during the early stages of the study. Kirill Korolev, Pankaj Mehta and the members of the Gore lab provided comments and advice on the manuscript. This research has been funded by NIH Pathways to Independence Award, NSF CAREER Award, Pew Biomedical Scholars Program and Alfred P. Sloan Foundation Fellowship.

**Author Contributions**

A.V. and J.G. designed research; A.V. performed research and analyzed data; A.V. and J.G. wrote the paper.

# SUPPLEMENTARY INFORMATION
## for
## THE STRENGTH OF GENETIC INTERACTIONS SCALES WEAKLY WITH THE MUTATIONAL EFFECTS

Andrea Velenich [1] and Jeff Gore [1]
[1] Department of Physics, Massachsetts Institute of Technology, Cambridge, MA 02139

**Figure S1**

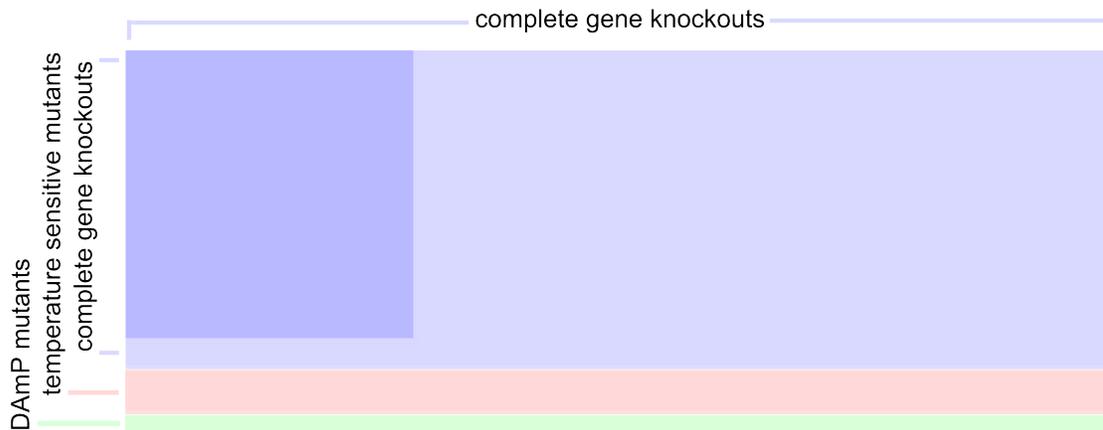

Figure S1 | The core of the DRYGIN data set is a 1712x3885 matrix whose entries are the growth rates of all possible double knockouts, obtained by crossing 1712 single knockouts of "query" genes with 3885 single knockouts of "array" genes. Since a few of the growth rates of the single knockouts mutants are reported as "NaN" (Not a Number), we restricted our analysis to a smaller 1570x3880 matrix, for which the growth rates of all single knockouts mutants are numerical values. Of the 1570 entries relative to query genes, 1287 are complete gene knockouts (blue rows); 191 are mutations which cause the gene product to misfold in a temperature-sensitive way (pink rows); 92 are DAmP mutants (Decreased Abundance by mRNA Perturbation) (green rows).

The darker blue square in the matrix represents double knockout mutants obtained from 1138 genes which appear both in the array and in the query subsets. For each pair of genes in this subset the double knockout mutants have been built twice, once as an "array x query" combination and once as a "query x array" combination. When both growth rates were numbers we used their average as the growth rate of the double knockout mutant. For about 5% of the pairs, one of the two combinations yielded "NaN", whereas the other one yielded a numerical growth rate; in those cases the numerical value was used as the growth rate of the double knockout mutant. For another 5% of the pairs, both growth rates were "NaN" and those pairs were dropped from the analysis. More generally, all double mutants with "NaN" and negative growth rates have not been considered during the analysis, whereas every mutant with a positive growth rate, even if very close to zero, has been considered.

**Figure S2**

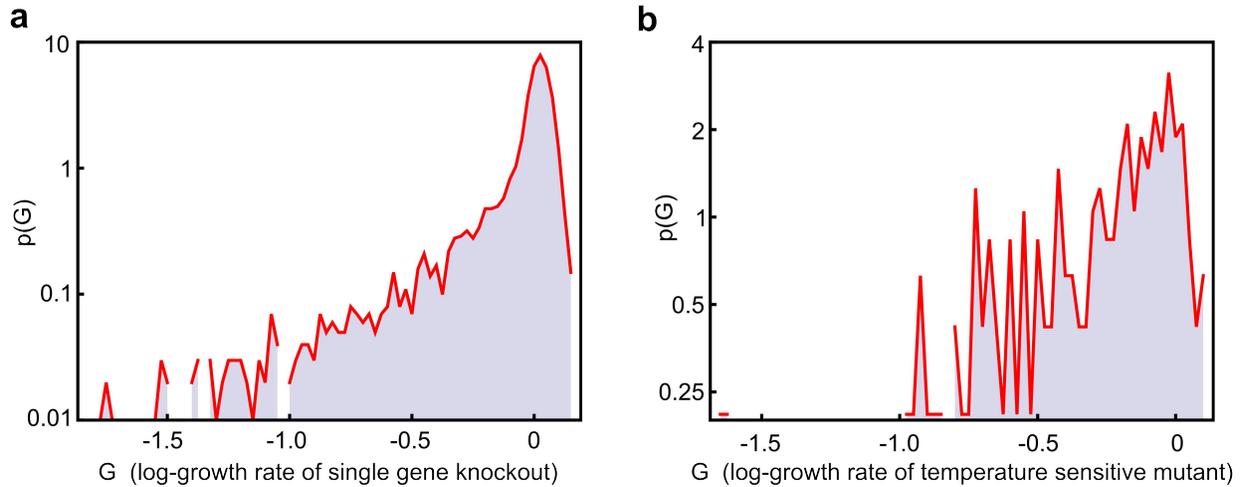

Figure S2 | Probability distributions p(G) of the log-growth effects for (a) all the single gene knockouts and (b) all the temperature sensitive mutants in the DRYGIN data set. The probability to observe a log-growth effect of G decreases approximately exponentially with |G|. Due to the scarcity of mutants with log-growth smaller than -1.0 and the consequent uncertainty in determining their statistical properties, the analysis in the main text considered only mutants for which G > -1.0 (corresponding to a growth rate of 0.5, relatively to the wild type).

**Figure S3**

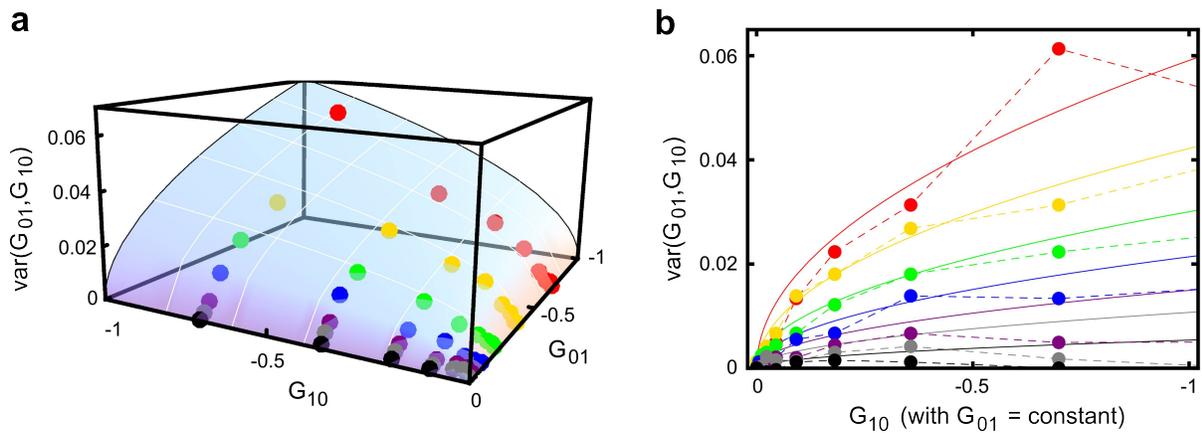

Figure S3 | The strength of epistatic interactions scales with the log-growth effects of the interacting knockouts. As in Figures 2a and 2b in the main text, each dot represents the variance of several thousand epistatic interactions binned according to the log-growth effects of the two single knockouts, $G_{01}$ and $G_{10}$. Here we show how the data can be alternatively fit by a simple power law. (a) The blue surface is the phenomenological fit: $\text{var}(G_{01}, G_{10}) = 0.071 \, \text{Sqrt}(|G_{01}||G_{10}|)$. (b) Slices of the plot in (a) for $G_{01}$ = constant. The dots are the same as in (a) and the solid lines represent the corresponding slice of the one-parameter fitting surface.

**Figure S4**

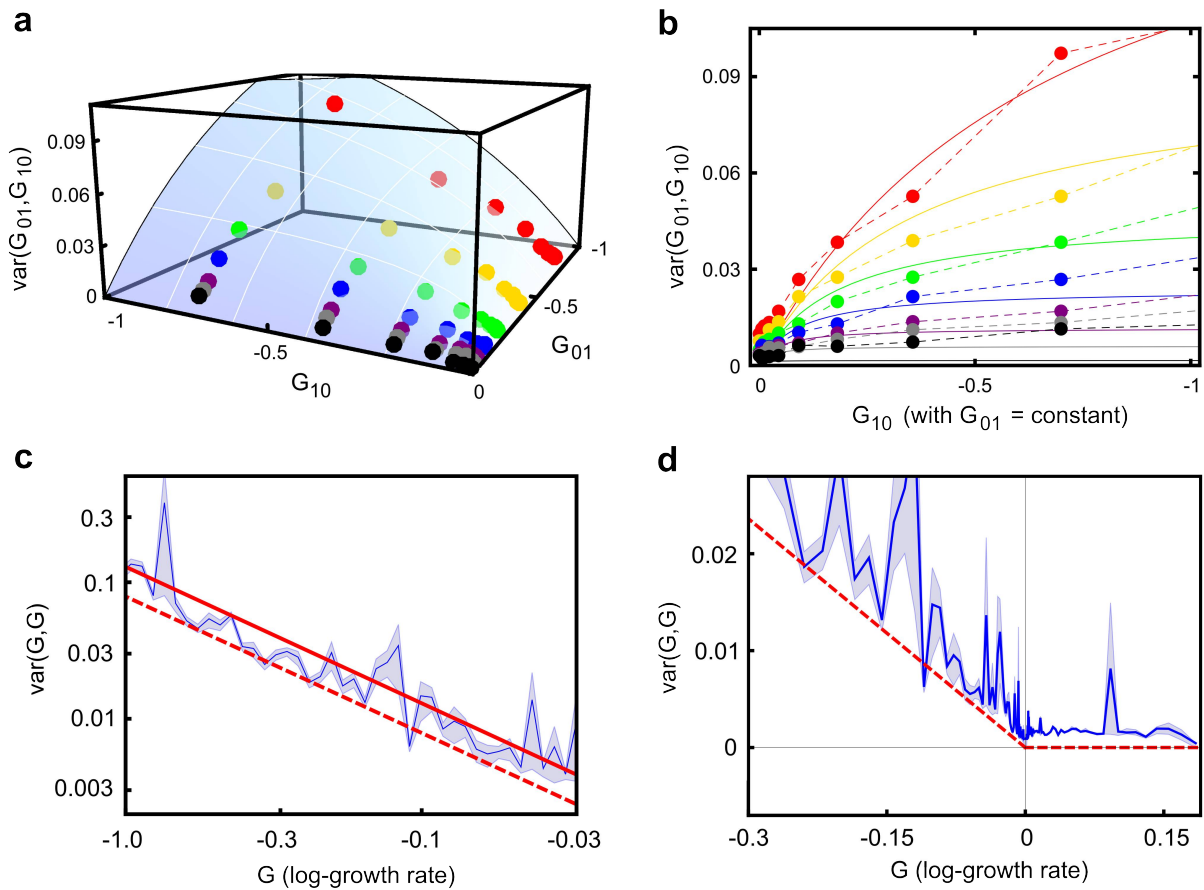

Figure S4 | The strength of epistatic interactions scales with the log-growth effects of the interacting knockouts. In contrast to Figures 2a, 2b, 2c and 2c Inset in the main text, here we plot the raw data, before subtracting the estimated contribution of the experimental noise. (a) Each dot represents the variance of several thousand epistatic interactions binned according to the log-growth effects of the two single knockouts, $G_{01}$ and $G_{10}$. The blue surface is the phenomenological fit: $\text{var}(G_{01}, G_{10}) = c \times 2 \, |G_{01}||G_{10}| / (|G_{01}| + |G_{10}|)$. In this case $c = 0.130$, whereas after subtraction of experimental noise $c = 0.079$. (b) Slices of the plot in (a) for $G_{01}$ =

constant. The dots are the same as in (a) and the solid lines represent the corresponding slice of the one-parameter fitting surface. (c) Diagonal slice of the plot in (a) with finer bins ($G_{01} = G_{10}$ within 20%, $G = \text{mean}(G_{01}, G_{10})$). The blue shaded area is the 25%-75% confidence interval computed by bootstrap; the solid red line ($\text{var}(G, G) = 0.130\ G$) is computed from the phenomenological model and the dashed red line ($\text{var}(G, G) = 0.079\ G$) is, for comparison, the corresponding line after subtracting experimental noise. (d) Diagonal slice of the plot in (a) showing both deleterious and beneficial knockouts. The dashed red lines are, for comparison, the fitting lines after subtracting experimental noise.

**Figure S5**

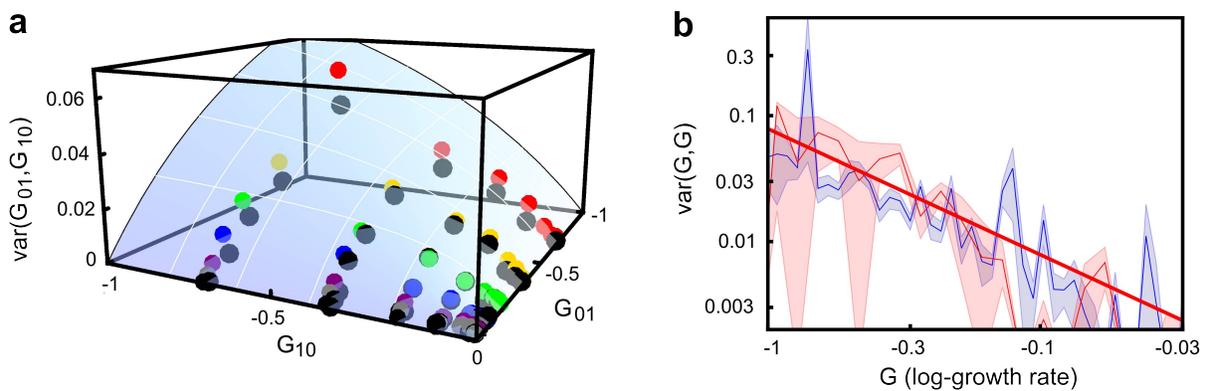

Figure S5 | DAmP "mutations" have similar epistatic interactions as entire gene knockouts. (a) Comparison between the epistasis observed in double gene knockout mutants (rainbow dots, same as in Fig. 2a) and the epistasis observed in mixed double mutants generated by combining a gene knockout and a DAmP (Decreased Abundance by mRNA Perturbation) perturbation of a different gene (black dots). (b) The red curve is the diagonal slice of the plot in (a) ($G_{01} = G_{10}$ within 20%, $G = \text{mean}(G_{01}, G_{10})$) and the red shaded area is the 25%-75% confidence interval for the mixed double mutants variance. For comparison it is superimposed to Figure 2c describing the variance for double gene knockouts (blue). As in Figure 2c the red line has equation $\text{var}(G, G) = 0.079\ G$.

**Figure S6**

The Gene Ontology (GO) database collects annotations of eukaryotic genes and it can be exploited to compare the traditional and the geometric definitions of epistasis in their ability to identify interacting pairs. Two genes are dubbed as GO-interacting if the number of GO terms they share is larger than some threshold (specifically, if two genes have n and m GO terms, we consider the set of all genetic pairs with n and m GO terms and dub as "GO-interacting" the 5% pairs sharing the highest number of GO terms in that set).

The traditional definition of epistasis identifies slightly more GO-interacting pairs: among the 10,000 most interacting pairs according to the traditional definition 1308 are GO-interacting, whereas among the top 10,000 most interacting pairs according to the geometric definition 1210 are GO-interacting (Fig. S6a). Both definitions agree in identifying 1010 GO-interacting pairs; 515 GO-interacting pairs would be expected by picking 10,000 random gene pairs. The 298 pairs which are identified as interacting only by the traditional definition tend to involve genes with small growth rate effects, whereas the 200 pairs which are identified as interacting only by the geometric definition tend to involve genes with large growth defects. Importantly, when doubling the threshold rank discriminating between interacting and non-interacting pairs, the geometric definition discovers almost all of the interactions discovered by the traditional definition with the original threshold. In contrast, even when doubling the threshold rank, the traditional definition misses many of the interactions discovered by the geometric definition with the original threshold (Fig. S6b). The GO enrichment of these neglected pairs indicates that the traditional definition of epistasis may completely overlook important interactions between genes with large growth rate defects.

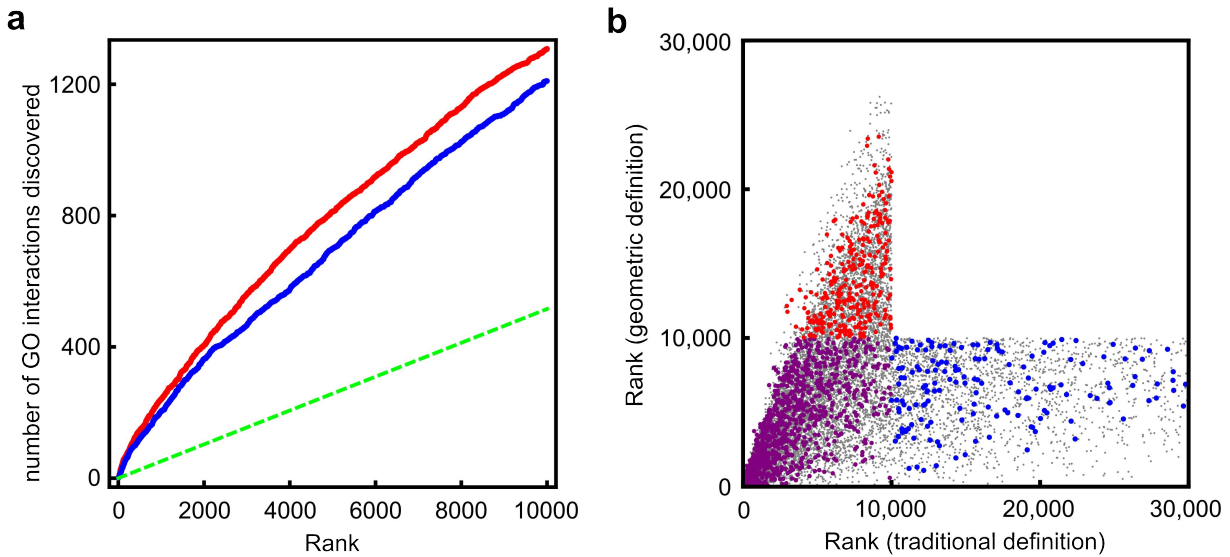

Figure S6 | Comparison between the traditional and the geometric definitions of epistasis. Each genetic pair is ranked according to the strength of the corresponding genetic interaction. (a) Number of Gene Ontology interactions discovered among the top-ranked pairs according to the geometric definition (blue), the traditional definition (red) and by selecting random pairs of genes (green dashed line). (b) Comparison of the ranks obtained by genetic pairs depending on how epistasis is defined. Each gray dot represent a genetic pair; larger dots represent interactions confirmed by the GO analysis (red: top-10,000 according to the traditional definition but not the geometric definition; blue: top-10,000 according to the geometric definition but not the traditional definition; purple: top-10,000 according to both definitions).

**Supplementary Text 1**

Rigorously introducing a new geometric model of epistasis compatible with all the features of the data is beyond the scope of this work. However it is worth providing some intuition for the experimentally observed dependence of the variance of epistasis:

$$\text{var}(G_{01}, G_{10}) = 2c \frac{|G_{01}||G_{10}|}{|G_{01}|+|G_{10}|}$$

Let us consider two genetic backgrounds, *WT* (wild type) and *A* (wild type plus mutation A). A second mutation, *B*, is known to have an effect $G_B$ in the wild type background. The unknown effect of mutation *B* in background *A* can be modeled as $G_B$ + N(0, $\sigma_B$=Sqrt(2c $G_B$)). The first term represents a deterministic "drift" which embodies our previous knowledge about the effect of mutation *B* in the wild type background, whereas the "diffusion" term N(0, $\sigma_B$) represents the uncertainty due to epistasis. By following the path *WT* → *A* → *AB*, we would then predict for the double mutant *AB* a log growth rate of $G_A$ + $G_B$ + N(0, $\sigma_B$).

We could also consider *WT* and *B* (wild type plus mutation B) as the two original backgrounds. Mutation *A* has then an effect $G_A$ in the wild type background and an effect $G_A$ + N(0, $\sigma_A$=Sqrt(2c $G_A$)) in background *B*. In this case, by following the path *WT* → *B* → *AB*, we would predict for the double mutant *AB* a log growth rate of $G_A$ + $G_B$ + N(0, $\sigma_A$).

Each of the two routes provides some *information* about the log growth rate of the double mutant *AB*. Importantly, the mutant *AB* is the same regardless of which route is chosen (*WT* → *A* → *AB* or *WT* → *B* → *AB*). Because of this constraint[49], the log growth rate of *AB* can be estimated as $G_{AB}$ = $G_A$ + $G_B$ + N(0, $\sigma_{AB}$), with N(0, $\sigma_{AB}$) proportional to N(0, $\sigma_A$) N(0, $\sigma_B$) and:

$$\sigma_{AB}^2 = \frac{\sigma_A^2 \sigma_B^2}{\sigma_A^2+\sigma_B^2} = 2c\frac{G_A G_B}{G_A+G_B} = \text{var}(G_A, G_B)$$

Notably, the model has only one free parameter, the "diffusion constant" c, conceptually analogous to a fitness landscape's roughness.

**Supplementary Text 2**

The Fisher's geometric model is a simple model of epistasis in which epistatic interactions emerge from geometry rather than being introduced *ad hoc* through random variables and noise. Phenotypes are described by *d* quantitative traits which can assume any real value and a particular phenotype is represented by a point *x* in the *d*-dimensional space. A fitness value is associated to each point in such space. Since most naturally occurring populations are believed to be in proximity of a local fitness optimum (which we set at *x*=0), a common choice for the fitness function is: f(*x*) = - ½ *x* · *x*, where the dot denotes the scalar product. A wild type organism with phenotype close to, but not exactly at the fitness optimum is represented by a vector $x_0 \neq 0$. It's fitness is then $f_0 = -\frac{1}{2} x_0 \cdot x_0$. A mutant is obtained by shifting the wild type phenotype in the *d*-dimensional space by a displacement d$x$. The fitness of the mutant is then $f(x_0+dx) = -\frac{1}{2}(x_0+dx) \cdot (x_0+dx) = -\frac{1}{2} x_0 \cdot x_0 - x_0 \cdot dx - \frac{1}{2} dx \cdot dx$, and the fitness effect of the mutation is df = $f(x_0+dx) - f(x_0) = -x_0 \cdot dx - \frac{1}{2} dx \cdot dx$. For small displacements df is approximately linear in |d$x$|, whereas for large displacements, or for displacements orthogonal to the fitness gradient, df is approximately proportional to |d$x$|$^2$.

Considering now two mutations and their combination in a double mutant:

$df_1 = -x_0 \cdot dx_1 - \frac{1}{2} dx_1 \cdot dx_1$

$df_2 = -x_0 \cdot dx_2 - \frac{1}{2} dx_2 \cdot dx_2$

$df_{12} = -x_0 \cdot (dx_1+dx_2) - \frac{1}{2}(dx_1+dx_2) \cdot (dx_1+dx_2)$

$E = df_{12} - df_1 - df_2 = -dx_1 \cdot dx_2$

Assuming that the displacements d$x$ are isotropic in the *d*-dimensional space, the average of their scalar product vanishes. Fisher's model predicts then that the mean epistasis is zero and that, on average, the effects of two mutations are additive. This observation points to the fact that the "fitness" values f in the Fisher's model are actually log-growth rates which, as we discussed in the main text, are on average additive variables.

In the Fisher's model, the standard deviation of epistatic interactions is:

$\sigma_E^2 = \text{mean}(E^2) - \text{mean}(E)^2 = \text{mean}((dx_1 \cdot dx_2)(dx_1 \cdot dx_2)) - 0 = |dx_1|^2 |dx_2|^2 \text{mean}(\cos^2(t))$

The means are computed over all possible orientations of the displacement vectors d$x_1$ and d$x_2$ in the *d*-dimensional space, with t representing the angle between d$x_1$ and d$x_2$. Mean($\cos^2(t)$) is a *d*-dependent numerical factor that can be computed analytically. Using the fact that df can be approximately proportional to |d$x$| or to |d$x$|$^2$, the characteristic magnitude of the epistatic interaction $\sigma_E(df_1, df_2)$ can be proportional to $df_1 df_2$ for small fitness effects and proportional to Sqrt($df_1 df_2$) for large fitness effects. In particular, when $df_1 = df_2 = df$, $\sigma_E(df, df)$ can be proportional to $df^2$ for small fitness effects and proportional to df for large fitness effects. As discussed in the main text, the experimentally observed dependence of the characteristic magnitude of the epistatic interactions with the log-growth effect of the mutations being combined is $\sigma_E(G, G) \propto$ Sqrt(G) and such dependence is weaker than the weakest dependence attainable in the Fisher's model.

**Supplementary Text 3**

For millions of mutants the DRYGIN data set provides a growth rate $g_{DRYGIN}$ and an estimate of the error on such growth rate $\sigma_{DRYGIN}$. Importantly, the value of "true" growth rate is not described by a normal distribution with mean $g_{DRYGIN}$ and standard deviation $\sigma_{DRYGIN}$. This is because the experimental error is estimated from only four independent growth measurements. If the "true" growth rate of a mutant is g and the randomness intrinsic in the experiment leads to an uncertainty $\sigma_g$, each independent growth measurement is a random number extracted from a normal distribution $N(g, \sigma_g)$ with mean g and standard deviation $\sigma_g$. Importantly, the value of the "true" growth rate, as estimated after four measurements, is distributed according to a student's t-distribution with three degrees of freedom, not according to the original normal distribution $N(g, \sigma_g)$. For a very large number of measurements the t-distribution converges to the original normal distribution and the distinction is not important. However, with four measurements, the t-distribution has tails decaying as $x^{-4}$ so that the difference between the estimated mean growth rate and the "true" growth rate can be much larger than what expected based upon the naïve assumption of an underlying normal distribution. Importantly, even if a t-distribution allows for larger fluctuations than a normal distribution, it's power-law tails are such that the variance remains finite, thus ensuring convergence for all the observables in the analysis carried out in the main text.

**Supplementary Reference**